\documentclass{article}

\usepackage[preprint]{neurips_2024}

\usepackage[utf8]{inputenc} 
\usepackage[T1]{fontenc}    
\usepackage{hyperref}       
\usepackage{url}            
\usepackage{booktabs}       
\usepackage{amsfonts}       
\usepackage{nicefrac}       
\usepackage{microtype}      
\usepackage{xcolor}         
\usepackage{amsmath}
\usepackage{float}
\usepackage{graphicx}

\title{Evaluating Voting Design Vulnerabilities for \\ Retroactive Funding}

\author{%
  Jay Yu \thanks{Funded in part through a research grant by the Optimism Foundation} \\
  Stanford University\\
  \texttt{jyu01@stanford.edu} \\
  \And
  Austin Bennett \\
  Stanford University\\
  \texttt{gaustinb@stanford.edu} \\
  \AND
  Billy Gao \\
  Stanford University\\
  \texttt{billygao@stanford.edu} \\
  \And
  Rebecca Joseph \\
  Stanford University\\
  \texttt{rjoseph5@stanford.edu} \\
}

\begin{document}

\maketitle

\begin{abstract}
Retroactive Public Goods Funding (RetroPGF) rewards blockchain projects based on proven impact rather than future promises. This paper reviews voting mechanisms for Optimism’s RetroPGF, where “badgeholders” allocate rewards to valuable projects. We explore Optimism's previous schemes for RetroPGF voting, including quadratic, mean, and median voting. We present a proof-based formal analysis for vulnerabilities in these voting schemes, empirically validate these vulnerabilities using voting simulations, and offer assessments and practical recommendations for future iterations of Optimism's system based on our findings.
\end{abstract}
\section{Introduction}\label{sec:intro}

\subsection{Background and Motivation}
Decentralized Autonomous Organizations (DAOs) have emerged as a novel form of organizational structure that leverages blockchain technology to enable transparent, automated decision-making processes \cite{hassan2021dao}. Many different kinds of DAOs exist today in the blockchain industry, including protocol DAOs such as Uniswap, Optimism, and Arbitrum, collector DAOs such as PleasrDAO and NounsDAO, and investor DAOs. As of writing (November 2024), DAOs today manage over 22 billion USD in treasury value, and with major DAOs such as Optimism holding over 3 billion USD in treasury value \cite{deepdao}.

Most DAOs today adopt some form of token-voting, where holders of a publicly-traded blockchain token, such as UNI, ARB, or OP are able to use their token holdings to cast votes on community proposals \cite{govbravo}. Token-based voting enables inclusive, community-driven participation, but practical challenges arise, including the risk of plutocratic capture by wealthy holders, low participation rates, and the need to balance efficiency with diverse interests \cite{austgen2023darkdao}. Additionally, blockchain’s pseudonymity introduces vulnerabilities like Sybil attacks and vote manipulation.

These real-world stakes and relative ease of changing governance processes within DAOs (compared with traditional corporate or political settings) have allowed DAOs to become a fertile testing ground for testing governance mechanism innovation \cite{hall2023op}.

\subsection{Optimism's RetroPGF}

Optimism's Retroactive Public Goods Funding (RetroPGF) represents an innovative program to sustaining public goods development in the blockchain ecosystem. Unlike traditional grant programs that fund projects based on prospective promises, RetroPGF rewards projects retroactively based on their demonstrated impact. This mechanism aligns incentives by ensuring that funding flows to projects that have already proven their value to the ecosystem \cite{opretropgf}.

The RetroPGF system operates through rounds of funding, where qualified voters (badgeholders) in Optimism's Citizen House evaluate and allocate rewards to projects that have contributed value to the Optimism ecosystem. The process begins with project teams submitting evidence of their contributions, followed by the careful selection of community members with demonstrated expertise as badgeholders. These badgeholders then engage in a thorough evaluation period, reviewing and assessing project contributions before their votes are aggregated to determine the final funding distribution.

This system presents unique challenges for voting mechanism design. The need to balance the expertise of specialized voters with broad community representation creates an inherent tension in the selection process. The system must prevent collusion while allowing voters to correctly express their project preferences. To address this, Optimism has experimented with different mechanisms in each fo the four rounds of prior RetroPGF funding \cite{opretropgf}.
\begin{itemize}
    \item Round 1 used a method of \textbf{Quadratic Voting} 
    \item Round 2 used a method of \textbf{Mean Voting}
    \item Round 3 and Round 4 used \textbf{Variants of Median Voting}
\end{itemize}

We will be exploring the voting designs and potential vulnerabilties of these various voting mechanisms in the following sections of this paper.

\subsection{Contributions}
This paper makes several contributions to the the analysis and development of Retroactive funding processes:

\begin{enumerate}
    \item We situate Optimism's RetroPGF within the broader context of social choice theory and offer a formal model for RetroPGF
    \item We provide proofs and analysis of  vulnerabilities within various RetroPGF processes
    \item We simulate RetroPGF voting processes to determine the practical impact of these theoretical vulnerabilities
    \item We provide specific recommendations for the implementation of voting mechanisms in future RetroPGF rounds and provide an open-source dashboard for practitioners to replicate and extend our findings.
\end{enumerate}

The remainder of this paper is organized as follows: Section~\ref{sec:litreview} provides an overarching literature review to situate RetroPGF within a broader context of social choice theory. Section~\ref{sec:methodology} outlines our methodology and introduces a formal model for our proof-based evaluation and simulation framework. Section~\ref{sec:proofs} details our proof-based analysis of vulnerabilities in the target voting mechanisms, and Section~\ref{sec:simulation} assesses their practical impact in a simulation context. Finally, we present our findings in Section~\ref{sec:results} and provide recommendations for further iterations of RetroPGF.
\section{Literature Review: Social Choice Theory and DAOs}\label{sec:litreview}

\subsection{Goals of Social Choice Theory}

Social choice theory examines how individual preferences can be aggregated into collective decisions while balancing fairness, efficiency, and strategic incentives \cite{sep-social-choice}. Its modern formulation traces back to Condorcet's work, which revealed that majority preferences could be cyclical: even with rational individual preferences, collective choices might yield intransitive outcomes where group preferences cycle between alternatives \cite{condorcet_condorcet_1994}. This observation, later known as Condorcet's Paradox, highlighted fundamental tensions in preference aggregation that continue to influence modern mechanism design, particularly the challenge of achieving \textit{strategyproof-ness}, where truthful participation is optimal regardless of others' actions.

Building on Condorcet's early work, two landmark results frame the modern study of social choice theory. Arrow’s Impossibility Theorem \cite{arrow1951social} proves no ranked voting system can simultaneously satisfy non-dictatorship, Pareto efficiency, and independence of irrelevant alternatives when there are three or more options. Gibbard-Satterthwaite’s theorem \cite{gibbard1973} extends this by showing all non-dictatorial voting schemes are susceptible to tactical voting unless restricted to binary choices. Together, these results establish fundamental limits on mechanism design: we cannot simultaneously achieve basic democratic principles and strategy-proofness except in highly restricted settings, shaping the problem space for modern DAO mechanism design.

\subsection{Theoretical Foundations for RetroPGF Mechanisms}

Optimism’s RetroPGF experiments operationalize social choice principles through three distinct voting mechanisms, quadratic voting in Round 1, mean voting in Round 2, and variants of median voting in Rounds 3 and 4. Here, we examine the core theoretical justifications for each voting mechanism.

\textbf{Quadratic voting}, used in RetroPGF Round 1, was developed by economists E. Glen Weyl and Steven P. Lalley \cite{lalley2018quadratic}, attempting to build on the earlier Vickrey-Clarke-Groves (VCG) mechanism to allow voters to express preference intensity \cite{posner2017}. In a token-based quadratic voting system, a voter's voting power is determined by the square root of the token holdings, allowing participants to express preference intensity while mitigating risk of power concentration among wealthy. Quadratic voting achieves allocative efficiency by imposing convex marginal costs on voting power, where the square root scaling ensures each additional unit of influence becomes exponentially more expensive. This convexity aligns individual expenditure with marginal utility, inducing voters to distribute tokens proportionally to their true valuations, thereby approximating welfare-maximizing equilibria observed in competitive markets \cite{posner2017}.

\textbf{Mean voting}, utilized in Round 2, has a less formally established framework than quadratic voting and median voting, but is conceptually very simple and builds on classic aggregation theory in statistics and economics\cite{galton1907vox}. Mean voting involves allocating funding according to the average of all votes received for each project, achieving representational accuracy by incorporating all votes equally into the final outcome, allowing full expression of preferences across the spectrum. However, this makes it vulnerable to manipulation through extreme values, just like linear voting. The mechanism optimizes for aggregate preference revelation, theoretically maximizing utilitarian welfare under assumptions of honest voting and normal distribution of preferences \cite{surowiecki2005wisdom}.

\textbf{Median voting}, adapted in Rounds 3 and 4 of Optimism's RetroPGF program, builds on Duncan Black's median voter theorem \cite{black1948rationale}, in which Black argues that the median allocation is robust against outlier values. The mechanism induces convergence toward centrist outcomes, reflecting the theoretical prediction that competing interests will gravitate toward the median voter's preference. However, its ordinal (vs cardinal) nature discards preference intensity data, creating openings for threshold-based attacks absent in interval-scale systems, such as phantom vote style attacks that we will describe in Section \ref{sec:proofs}.

\subsection{Cross-Mechanism Comparison with DAOs}

Traditionally, each of the above mechanisms have been analyzed and theorized in isolation, often in the context of existing political and economic systems. However, the rapid ascent of DAOs provide a new paradigm for studying multiple voting mechanism innovations in conjunction with one another in a setting with real-world stakes \cite{hall2023op}. In particular, Optimism's sequential adoption of multiple mechanisms, from quadratic (Round 1), mean (Round 2), and median (Rounds 3-4) voting provide a valuable real-world testbed for comparative vulnerability analysis under shared model constraints, which are described in Section \ref{sec:methodology}. 

In our work, we present a cross-mechanism comparison through a dual-approach, creating formalized proofs of different vulnerabilities and attack strategies on each of the target mechanisms, and a simulation framework that seeks to quantify and compare the resulting impacts of each minimum viable attack strategy on the various voting mechanisms. Our work thus provides a common baseline and lens of analysis to assess the comparative strengths and weaknesses of these various mechanisms from social choice theory, while also giving recommendations to DAO practitioners on how to design a more robust voting system for RetroPGF.
\section{Methodology}\label{sec:methodology}

As mentioned in Section~\ref{sec:intro}, Optimism has previously conducted RetroPGF using three different mechanisms across their four rounds: quadratic voting, mean voting, and median voting. Our approach will focus on both a proof-based theoretical analysis of these voting mechanisms to expose potential vulnerabilities in each of these models in Section~\ref{sec:proofs}. We then empirically verify and assess the impact of these vulnerabilities in simulated voting rounds in Section~\ref{sec:simulation} and discuss our results in Section~\ref{sec:results}. In this section, we will outline the formal model of Optimism's RetroPGF setup that we will analyze, both in our formal analysis and our simulations.

\subsection{Optimism RetroPGF Governance Assumptions}

In their RetroPGF program, Optimism incentivizes development by providing retroactive funding to projects based on their success, which is determined by a group of pre-selected badge holders. To model this process, in this paper, we apply the following assumption model:

\begin{enumerate}
   \item There is a fixed number of voters in our system.
   \item A voter cannot split their funds into multiple voting wallets.
   \item There is a fixed number of projects where voters can allocate funds. 
   \item Voters all have a set number of tokens to allocate.
   \item Rational voters will maximize their utility -- a function of their  allocations to projects.
\end{enumerate}

\subsection{Model Specification}
Consider a single-shot Optimism RetroPGF round with $n \in \mathbb{Z}^+$ voters and a total funding allocation of $T \in \mathbb{R}^+$ tokens to be evenly divided among them. This yields a voting power of $\frac{T}{n}$ for each voter. We assume that there are no costs associated with casting a vote. Thus, we define the following model to represent a funding round:

\begin{itemize}
    \item The number of voters within a retroactive funding round: $N \in \mathbb{Z}^+$
    \item The number of projects available for funding: $P \in \mathbb{Z}^+$
    \item The number of tokens to be distributed: $T \in \mathbb{R}^+$
    \item The Allocation Matrix, $A \in \mathbb{R}^{NxP}$, where $A_{ij}$ is equal to voter $i$'s allocation to project $j$.
\end{itemize} 

\textit{Subject to the Following Model Variable Constraints}
\begin{enumerate}
    \item All voters have a maximum of $\frac{T}{N}$ votes: $\sum_j^P A_{ij} \leq \frac{T}{N}, \forall i \in \left[ 1, ..., N \right]$
    \item Vote amounts are non-negative: $A \geq 0$
\end{enumerate}

\subsection{A Simplified Model of Aggregate Voting Mechanisms}

\subsubsection{Mean Voting}
Given the allocation matrix $A \in \mathbb{R}^{N \times P}$, where $A_{ij}$ represents the allocation of tokens from voter $i$ to project $j$, the total allocation for each project $j$ is calculated as:

$$T_j = \sum_{i=1}^N A_{ij}$$

The allocation is given by:

\[
P_j = \frac{T_j}{\sum_{j=1}^P T_j}.
\]
\[\text{ Allocation = } T \cdot P_j \]

\subsubsection{Median Voting}

Given the allocation matrix $A \in \mathbb{R}^{N \times P}$, where $A_{ij}$ represents the allocation of tokens from voter $i$ to project $j$, the median allocation for each project $j$ is calculated as:

\[
M_j = \text{median} \{ A_{ij} \mid A_{ij}, \, i = 1, \dots, N \},
\]

The allocation is given by:

\[
P_j = \frac{M_j}{\sum_{j=1}^P M_j},
\] 
\[\text{ Allocation = } T \cdot P_j \]

\subsubsection{Quadratic Voting}
Given the allocation matrix $A \in \mathbb{R}^{N \times P}$, where $A_{ij}$ represents the allocation of tokens from voter $i$ to project $j$, the quadratic voting allocation for each project $j$ is calculated as:

\[
Q_j = \sum_{i=1}^N \sqrt{A_{ij}},
\]

The allocation is given by:
\[
P_j = \frac{Q_j}{\sum_{j=1}^P Q_j}.
\] 
\[\text{ Allocation = } T \cdot P_j \]

\section{Voting Mechanism Analysis}\label{sec:proofs}

\subsection{Quadratic Voting}

First proposed by Lalley and Weyl \cite{lalley2018quadratic}, a quadratic voting mechanism reweighs votes by valuing them as the square roots of the allocation. This has many advantages with respect to economic optimality and decentralization; however, in many blockchain systems it is difficult to implement due to the lack of identity verification and the possibility of Sybil attacks, where an adversary could split their tokens among an arbitrarily large number of wallets and gain an outsize influence on the network.

In the context of Optimism's RetroPGF program, however, this is not a concern as voters cannot split their token votes among some large number of wallets. Scenarios resembling Sybil attacks could still persist should multiple voters decide to collude. Specifically, two voters with strong preferences for one project who collude can artificially increase their voting power by a factor $\sqrt{2}$. To see this, consider the optimization problem.

\newtheorem{claim}{Remark}

\begin{claim}{The Quadratic Collusion Attack -- For two colluding voters, who care solely about the success of projects $p$ and $q$, respectively, optimal collusion for both parties under Quadratic Voting allows them to each artificially increase their voting power by a factor of $\sqrt{2}$.}
\end{claim}

Consider two colluding voters have a total of $T$ tokens who can contribute $(p_1$, $q_1)$ and $(p_2$, $q_2)$ to projects $p$ and $q$, respectively, yielding total contributions of $P = p_1 + p_2$ and $Q = q_1 + q_2$. Additionally, assume that these voters only care about the number of funds allocated to projects $p$ and $q$, respectively. Holding all other allocations by voters constant, this yields the simplified utility functions:

\begin{itemize}
    \item Voter 1 Utility: $u_1(p_1$, $q_1) = \sqrt{p_1} + \sqrt{p_2}$
    \item Voter 2 Utility: $u_2(p_2$, $q_2) = \sqrt{q_1} + \sqrt{q_2}$
\end{itemize}

To find optimal allocations for both colluding parties, we know that neither will voter will allow the other project to amass more of their collective resources, as this would be suboptimal since voting power increases with votes allocated. Thus, we know that $P = Q = T$. 

Now, the task is to find the optimal allocation among $p_1$, $p_2$, $q_1$, and $q_2$. Our constraints only restrict the relationship between $p_1$ and $p_2$ as well as  $q_1$ and $q_2$, so we can treat these as separate optimization problems and solve for the optimal value for our decision variables.

Starting with $p_1$ and $p_2$, we have the following optimization problem:

$$\max_{p_1, p_2} \sqrt{p_1} + \sqrt{p_2}$$
Subject to:
\begin{enumerate} \vspace{-0.2cm}
    \item $p_1 + p_2 = P$
    \item $p_1 \geq 0$
    \item $p_2 \geq 0$
\end{enumerate} 

The second and third constraints are satisfied by our model definition. Additionally, by the first constraint, we can rewrite $p_2 = P - p_1$. Substituting this into our optimization problem, we obtain a single-variable objective function to maximize. Taking the first order conditions, we can solve for the optimal value as $p_1$ as follows: \vspace{-0.2cm}

\begin{align*} 
    \max_{p_1} \sqrt{p_1} + \sqrt{P - p_1} 
    & \implies \frac{\partial}{\partial p_1}(\sqrt{p_1} + \sqrt{P - p_1}) = 0 \\ \\
    & \implies (\frac{1}{2\sqrt{p_1}} - \frac{1}{2\sqrt{P - p_1}}) = 0 \\ \\
    & \implies \frac{1}{2\sqrt{p_1}} = \frac{1}{2\sqrt{P - p_1}} \\ \\
    & \implies \sqrt{p_1} = \sqrt{P - p_1} \\ \\
    & \implies p_1 = P - p_1 \\ \\
    & \implies 2p_1 = P \\ \\
    & \implies p_1 = \frac{P}{2} \\ \\
\end{align*}

From here, we see that the optimal allocation for $(p_1^*, p_2^*) = (\frac{P}{2}, \frac{P}{2})$, yielding the optimal utility:

\[u_1^*(p_1^*, p_2^*) = 2\sqrt{\frac{P}{2}} = \sqrt{2} \cdot \sqrt{P} = \sqrt{2} \cdot \sqrt{T}\]

Without loss of generality, the same argument can be applied in solving for the optimal $(q_1^*, q_2^*) = (\frac{Q}{2}, \frac{Q}{2})$, yielding the optimal utility:

$$u_2^*(q_1^*, q_2^*) = 2\sqrt{\frac{Q}{2}} = \sqrt{2} \cdot \sqrt{Q} = \sqrt{2} \cdot \sqrt{T}$$

In both cases, we see that the colluding parties, each of whom in theory have a voting power of $\sqrt{T}$, can multiply their voting power by a factor of $\sqrt{2}$ by agreeing to use half of their tokens to vote for the other project. $\square$ 

\subsection{Mean Voting}
Mean voting is a mechanism were we allocate funding according to the mean of the vote amounts for a project. Here, we show the importance of including all wallets under the aggregation process, including those with a vote of zero tokens.

\begin{claim}{When all vote allocations are included, then Mean Voting is a Linear Voting Mechanism}
\end{claim}

When all votes are included, notice that for a project, $j$ in our model, $j \in \left[ 1, ..., P \right]$, we can represent its allocation total as:

\[\text{Project j's Mean Votes Received} = \frac{1}{N}\sum_{i=1}^N A_{ij}\]

Since $N$ is a constant, we see that mean allocation has a linear relationship to the number of tokens allocated to the project. This is proportionally the same as Linear Voting; and, therefore, offers no strategic advantage in resources allocation. It may, however, yield some advantages with respect to ease of determining token allocation. For example, mean voting could allow groups to make decisions about the allocation proportion from a larger pool in an easy-to-view way for voters. 

\begin{claim}{The Mean Phantom Vote Attack -- Where $n$ is the number of non-zero wallets, Mean Voting without zero-value allocations included can lead k adversaries to artificially reduce the vote total by up to $\frac{n}{n+k}$, where $k \leq N - n, k \in \mathbb{Z}^+$.}
\end{claim}

To see this, choose some arbitrarily project $j$ with $n$ non-zero allocations and the allocation vector $a_j = (a_{1j}, ..., a_{nj})$. Notice that this is not the same as the allocation matrix, as this is simply the non-zero subset of the column associated with allocations for a project, $j$. Thus, the mean voting power prior to an attack simply takes the form:

\[\text{Project $j$'s Mean Voting Power before Attack} = \frac{1}{n}\sum_{i=1}^n a_{ij}\]

Now, say some, $k$, adversaries don't like this project and want to decrease its mean vote total. This can be for a variety of reasons, though most commonly to reserve funds for another project.

The adversaries can each choose some arbitrarily small $\epsilon_k > 0$ for them to commit to a project, j. This yields, with a slight abuse of notation, the following mean voting power after an attack:

\[\text{Project $j$'s Mean Voting Power after Attack} = \lim_{\epsilon_i \to 0^+}\frac{1}{n+k}(\sum_{i=1}^n a_{ij} + \sum_{i=1}^k\epsilon_i) = \frac{1}{n+k}\sum_{i=1}^n a_{ij}\]

From this, we see that the relative mean voting power of a project after the attack compared to before the attack can be simplified as follows:

\[\frac{\text{Project $j$'s Mean Voting Power after Attack}}{\text{Project $j$'s Mean Voting Power before Attack}} = \frac{\frac{1}{n+k}\sum_{i=1}^n a_{ij}}{\frac{1}{n}\sum_{i=1}^n a_{ij}} = \frac{n}{n+k} \square\]

This problem arises from the number of voters which we count in the mean vote total. Since an adversary can contribute effectively nothing, they can decrease the average number of vote allocation. If we count all vote allocations, even with zero value, then we have the constant number of participants, $N$, yielding a scaling term of $\frac{1}{N}$ for all vote allocation totals. This yields the linear system described in Remark 1.

\subsection{Median Voting}
Median voting is a mechanism where we allocate funding according to the median of the vote counts for a project. Here, we show how the phantom vote attack may be more dangerous for median voting, as we cannot rely on including all zero votes as we did with mean voting.

\begin{claim}{The Median Phantom Vote Attack -- Where $n$ is the number of non-zero wallets, let $a_j^{(i)}$ be the $ith$ order statistic for $i \in \left[1, ..., n\right]$. If we have a median allocation $a_{m}$, where $m$ is the largest index such that $a_{m} \leq median(a)$, then Median Voting without zero-value allocations included can lead k adversaries to artificially decrease the voting power to where it is bounded above by $a_j^{(m-\lceil\frac{k}{2}\rceil+1)}$.}
\end{claim}

To see this, invoke a similar strategy as described in the "mean phantom vote" section. More specifically, each adversary, k, can each choose some corresponding arbitrarily small $\epsilon_k > 0$ for them to commit to a project, j, such that $\epsilon_k < a_1$ prior to the attack. \\

Because the calculation of median depends on the number of elements, we now consider four cases: \\

\underline{Case: n is even, k is even} \\
In this case, we see $n + k$ is even. Prior to the attack, the median value is equal to $\frac{a_j^{(m)} + a_j^{(m+1)}}{2}$. Because we now have $k$ elements below our median, by the definition of median we see that our median value shifts down $\frac{k}{2}$ order statistics and is now $\frac{a_j^{(m-\frac{k}{2})} + a_j^{(m-\frac{k}{2}+1)}}{2}$, as we still have an even number of allocations and need to take the average of the allocations on either side. \\

This is bounded above by $a_j^{(m-\lceil\frac{k}{2}\rceil+1)}$, as we have:

\[a_j^{(m-\frac{k}{2})} \leq a_j^{(m-\lceil\frac{k}{2}\rceil+1)} \implies \frac{a_j^{(m-\frac{k}{2})}}{2} \leq \frac{a_j^{(m-\lceil\frac{k}{2}\rceil+1)}}{2} \implies \frac{a_j^{(m-\frac{k}{2})} + a_j^{(m-\lceil\frac{k}{2}\rceil+1)}}{2} \leq a_j^{(m-\lceil\frac{k}{2}\rceil+1)}\] \\

\underline{Case: n is even, k is odd} \\
In this case, we see $n + k$ is odd. Prior to the attack, the median value is equal to $\frac{a_j^{(m)} + a_j^{(m+1)}}{2}$. Because we now have $k$ elements below our median, by the definition of median we see that our median value shifts down $\frac{k-1}{2}$ order statistics from $a_j^{m}$ and is now $a_j^{(m-\frac{k-1}{2})}$.  \\

This is bounded above by $a_j^{(m-\lceil\frac{k}{2}\rceil+1)}$. \\ \\

\underline{Case: n is odd, k is even} \\
In this case, we see $n + k$ is odd. Because we now have $k$ elements below our median, by the definition of median we see that our median value, $a_j^{(m)}$, shifts down $\frac{k}{2}$ order statistics and is now $a_j^{(m-\frac{k}{2})}$. \\

This is bounded above by $a_j^{(m-\lceil\frac{k}{2}\rceil+1)}$. \\

\underline{Case: n is odd, k is odd} \\
In this case, we see $n + k$ is even. Because we now have $k$ elements below our median, by the definition of median we see that our median value, $a_j^{(m)}$, shifts down $\frac{k-1}{2}$ order statistics and is now $\frac{a_j^{(m-\frac{k-1}{2})} + a_j^{(m-\frac{k-1}{2} + 1)}}{2}$. \\

This is bounded above by $a_j^{(m-\lceil\frac{k}{2}\rceil+1)}$. \\

Thus, we see this bound holds in all cases so the median phantom vote attack is proven. While perhaps more difficult to quantify, the phantom vote attack may be more dangerous in this setting than mean voting. This is because if a large gap between two allocations is identified, it could be targeted by a sufficient number of adversaries to exploit, causing a much larger jump downward than we see with mean voting. For this reason, we may wish to implement alternative measures, such as a minimum allocation threshold in this case.

\section{Voting Mechnanism Simulations}\label{sec:simulation}

\subsection{Experimental Design} \label{sec:simulation-exp-design}
This study uses a simulation-based framework to evaluate the performance and robustness of various voting mechanisms in Optimism's RetroPGF process \footnote{The complete codebase used for the simulation can be found at \url{https://github.com/Stanford-Blockchain-Club/opt-voting}}. Our simulation runs 10,000 iterations with 133 voters and 374 projects, parameters chosen to match the scale of real-world RetroPGF rounds \cite{op_rpgf_project_stats}.

For each voting mechanism (Quadratic, Mean, and Median), our experimental evaluation focuses on two primary attack vectors:
\begin{enumerate}
    \item \textbf{Voter Attacks:} Scenarios where groups of voters coordinate to manipulate outcomes by strategically concentrating their voting power on specific projects.
    \item \textbf{Project Attacks:} Scenarios where multiple projects collude by encouraging their supporters to split votes among the colluding group, effectively amplifying their collective allocation.
\end{enumerate}

In addition to these two attack scenarios, we also model a baseline allocation for each target voting mechanism. We measure the voting mechanisms resilience through a pairwise comparison between the baseline allocation and post-attack allocations, defined in Section~\ref{sec:sim-eval-metric}, to compute a Pairwise Manipulation Score that we compare across different attack scenarios.

\subsection{Preference Construction}

Our simulation results are based on a voter preference matrix, denoted $M$, representing the relative preferences of the voters distributed throughout all the projects, roughly corresponding to the Allocation Matrix $A$ in Section~\ref{sec:methodology}. Each element of the matrix, $M_{v,p}$ represents the preference of voter $v$ for project $p$, normalized such that the preferences for each voter sum to 1. For our simulation, we construct $M$ voter by voter such that:
\begin{equation}
M_{v,p} = \frac{X_{v,p}}{\sum_p X_{v,p}} \quad \forall v
\end{equation}\noindent
where $X_{v,p}$ is sampled from a Pareto distribution:\vspace{0cm}
\begin{equation}
X_{v,p} \sim \text{Pareto}(\alpha=2.5) \quad \forall v,p
\end{equation}\noindent
This construction ensures:\vspace{0cm}
\begin{equation}
M_{v,p} \in [0,1] \quad \forall v,p
\end{equation}\vspace{0cm}
\begin{equation}
\sum_p M_{v,p} = 1 \quad \forall v
\end{equation}

We generate $M$ using a Pareto distribution to better model empirical observations that each individual voter may only have time to thoroughly evaluate a small subset of total projects, as well as herding and other naturally occurring centralizing vectors for preference distribution \cite{sharma2023unpacking}. The shape parameter $\alpha = 2.5$ models moderate preference concentration while maintaining finite variance in the distribution, allowing for moderately heavy tails while also accounting for preference concentration to better match empirical observations for RetroPGF rounds. Our simulation framework also supports alternative distributions of $X_{v, p}$ such as sampling from a uniform or Gaussian distribution, or directly importing voter preference data.

In addition to matrix $M$, our simulation framework also accounts for a weight vector $W$, where each $w_v \in W$ accounts for voter $v$'s vote weights. As per our Optimism RPGF model specification in Section~\ref{sec:methodology}, in our experiment $\forall v(w_v = 1)$. Then, voter weights are normalized such that $\sum_v w_v = c$, where $c$ is a normalization constant chosen large enough for floating point numerical stability (eg. 1000).


\subsection{Voting Mechanisms}

As outlined in Section ~\ref{sec:simulation-exp-design}, we examine how three different voting mechanisms (Quadratic, Mean, and Median) that are used in prior RetroPGF rounds behave under Voter Attack and Project Attack scenarios. These two attack scenarios represent two different sets of incentives for attacking, one from a voter perspective, wanting their chosen projects to win, and the other from a project perspective. We also add a baseline mechanism that simply sums up voter preferences to act as a control. Table~\ref{tab:voting-mechanisms} provides an overview of the different strategies in each of the voting mechanisms and attack scenarios.

\begin{table}[H]
\caption{Voting Mechanisms and Attack Scenarios}
\label{tab:voting-mechanisms}
\begin{center}
\begingroup
\renewcommand{\arraystretch}{1.5}
\begin{tabular}{|l|p{3.8cm}|p{3.8cm}|p{3.8cm}|}
\hline
\textbf{Mechanism} & \textbf{Baseline} & \textbf{Voter Attack} & \textbf{Project Attack} \\ \hline

Baseline & 
$R_p = \sum_v M_{v,p} \cdot w_v$ &
N/A &
N/A \\ \hline

Quadratic & 
$R_p = \sum_v \sqrt{M_{v,p} \cdot w_v}$ & 
Colluding pairs split $\sqrt{0.5w_v}$ between top preferences &
Projects pool supporters to split votes among colluding group \\ \hline

Mean & 
$R_p = \frac{1}{|V|}\sum_v M_{v,p} \cdot w_v$ & 
Attacker assigns $\varepsilon$ to all except max preference & 
Project supporters minimize votes to other projects ($\varepsilon$) \\ \hline

Median & 
$R_p = \text{median}_v(M_{v,p} \cdot w_v)$ &
Similar to mean attack with $\varepsilon$ strategy &
Similar to mean attack with coordinated $\varepsilon$ assignments \\
\hline

\end{tabular}
\endgroup
\end{center}
\small\textit{Note:} $M_{v,p}$ represents normalized preferences where $\sum_p M_{v,p} = 1$ for each voter $v$. $w_v$ is voter $v$'s voting power. $\varepsilon$ represents a small positive constant (0.01).
\end{table}

\subsubsection{Baseline Mechanism}
The baseline mechanism directly aggregates normalized voter preferences, providing a benchmark for comparing other mechanisms:

\begin{equation}
R_p = \sum_v M_{v,p} \cdot w_v
\end{equation}

where $M_{v,p}$ represents normalized preferences such that $\sum_p M_{v,p} = 1$ for each voter, serving just as a reference point for measuring manipulation in other mechanisms.

\subsubsection{Quadratic Voting}
Quadratic voting applies a square root transformation to voting power to reduce the impact of vote concentration:

\begin{equation}
R_p = \sum_v \sqrt{M_{v,p} \cdot w_v}
\end{equation}

Under the voter attack scenario, a pair of colluding voters ($v_1, v_2 \in C$) coordinate by splitting their voting power between their highest-ranked preferences:

\begin{equation}
R_p = \begin{cases} 
\sqrt{0.5w_v} & \text{if } p \in \text{top}_2(M_{v,p}), v \in C \\
\sqrt{M_{v,p} \cdot w_v} & \text{otherwise}
\end{cases}
\end{equation}

Project attacks occur when an attack project ($P_c$) coordinate their supporters ($V_p$) to split votes among the colluding group:

\begin{equation}
R_p = \begin{cases}
\sqrt{0.5w_v} & \text{if } p \in P_c, v \in V_p \\
\sqrt{M_{v,p} \cdot w_v} & \text{otherwise}
\end{cases}
\end{equation}

\subsubsection{Mean Voting}
Mean voting averages normalized allocations across all voters:

\begin{equation}
R_p = \frac{1}{|V|}\sum_v M_{v,p} \cdot w_v
\end{equation}

This mechanism adapts phantom voting techniques from our theoretical framework for a fixed voter set, as required in our model in Section~\ref{sec:methodology}. Under voter attacks, an attacker allocates minimum votes ($\varepsilon$) to all projects except their target to manipulate the mean:

\begin{equation}
R_p = \frac{1}{|V|}\sum_v \begin{cases}
w_v - (|P|-1)\varepsilon & \text{if } p = p^*_v \\
\varepsilon & \text{otherwise}
\end{cases}
\end{equation}

Project attacks follow a similar pattern but coordinate across supporter groups:

\begin{equation}
R_p = \frac{1}{|V|}\sum_v \begin{cases}
w_v - (|P|-1)\varepsilon & \text{if } p \in P_c, v \in V_p \\
\varepsilon & \text{otherwise}
\end{cases}
\end{equation}

\subsubsection{Median Voting}
As in our theoretical framework, median voting follows similar phantom attack patterns to mean voting but uses the median to reduce sensitivity to extreme allocations. We again use the $\epsilon$ pattern to model these here. For voter attacks:
\begin{equation}
R_p = \text{median}_v \begin{cases}
w_v - (|P|-1)\varepsilon & \text{if } p = p^*_v \\
\varepsilon & \text{otherwise}

\end{cases}
\end{equation}
Project attacks similarly coordinate across supporter groups:
\begin{equation}
R_p = \text{median}_v \begin{cases}
w_v - (|P|-1)\varepsilon & \text{if } p \in P_c, v \in V_p \\
\varepsilon & \text{otherwise}

\end{cases}
\end{equation}

For each mechanism, we implement minimum viable attacks — the simplest attack patterns that demonstrate the mechanism vulnerabilities that consist of having a minimum amount of voters or projects participating in an attack type.

\subsection{Evaluation Metric} \label{sec:sim-eval-metric}

To quantify the resilience of voting mechanisms against manipulation, we introduce a Pairwise Manipulation Score (PMS), a modification of Mean Squared Error (MSE) adapted specifically for voting distribution analysis. While MSE measures absolute differences between distributions, PMS is designed to capture relative changes in voting power distribution.

For vote distributions $v_1, v_2$ over $n$ projects, we first normalize to percentage allocations:

\begin{equation}
    p_i^k = \frac{v_k[i]}{\sum_{j=1}^n v_k[j]} \times 100
\end{equation}

where $p_i^k$ represents the percentage of votes allocated to project $i$ in distribution $k$. The Pairwise Manipulation Score is then defined as:

\begin{equation}
    \text{PMS}(v_1,v_2) = \frac{\sum_{i=1}^n (p_i^1 - p_i^2)^2}{\sum_{i=1}^n (p_i^1)^2} \times 100
\end{equation}

Unlike traditional MSE which takes the form $\frac{1}{n}\sum_{i=1}^n (y_{\text{true}} - y_{\text{pred}})^2$, our metric creates two key modifications, with percentage normalization to ensure scale invariance across different voting mechanisms and total vote counts, and division over the sum of squared baseline percentages to measure relative rather than absolute changes. For each voting mechanism $M$ and attack scenario $A$, we can thus compute:
\begin{equation}
    \text{Manipulation}(M,A) = \text{PMS}(\text{BaselineResults}(M), \text{AttackResults}(M,A))
\end{equation}

A PMS of 0 indicates identical distributions, while higher scores represent increasing levels of vote distribution distortion, quantitively comparing the resilience of different voting mechanisms in the different attack scenarios.

\section{Results}\label{sec:results}

\subsection{Theoretical Evaluation}

Drawing from the proofs of our voting model in Section ~\ref{sec:proofs}, we have a few observations: \\

\underline{Observation 1: Median Voting is More Vulnerable to Large-Scale Attacks than Mean Voting} \\ \\
While both measures attempt to find some notion of a "middle" allocation, notice that the upper bounds of our Mean Phantom Vote Attack and Median Phantom Vote Attack are quite different. The former relies on a normalized sum, while the latter relies on an ordering of the allocations.

For this reason, each allocation will have an equal impact on the value of Mean Voting, while only the center allocations impact the value of a Median Voting, notwithstanding the actual order of the allocations themselves. Thus, an attacker could cause much massive swings within a median value, especially if there are large gaps in value between the allocations.

To get an intuition for why this difference matters, consider the following set of allocations for a project: $\left[0, 0, 0, 0, 100, 100, 100, 100\right]$. Currently, both mechanisms yield a value of $50$.

If an adversary contributes $0$ to both projects, we now have $\left[0, 0, 0, 0, 100, 100, 100, 100\right]$, yielding: \\

Mean Voting = $\frac{1}{9}(5 \cdot 0 + 4\cdot100) = 44.\overline{4} $ \\ \\
Median Voting = 5th Ranked Value = 0\\

This is an extreme example, as median is unlikely to be swayed this much by a single vote; but in as we've shown in $\text{Section } 4.2$, many adversaries can attack to bring down the median. Additionally, we can't simply count zero-valued votes as we did in Mean Voting, as this would also bring down the median to a point where it could wash out all real allocations.

There have been some clever workarounds proposed to address other issues in Median Voting, such as the use of Capped Median Rule and Moving Phantoms Rules that could provide some sort of stopgap against this attack due to making this drop off less severe. Additionally, raising the threshold of a minimum allocation and other floors and ceilings within mechanism design could make this sort of attack prohibitively expensive. We discuss these additional considerations in Section ~\ref{sec:further}. \\

\underline{Observation 2: Quadratic Voting Exploitation Occurs with Repeated Voters} \\ \\
While the prospect of increasing vote power by a factor $\sqrt{2}$ is quite substantial, this only works so long as two delegates have an incentive to collude. 

If the same delegates are consistently used for every vote, over time the benefits of collusion could outweigh the benefits of betraying the effort. Its possible that this repeated collusion never reaches a Nash Equilibrium; but if the delegates are frequently switched, it becomes substantially less likely as the delegates would almost certainly value future collusion less.

At a high-level, if a colluder is going to contribute half of their tokens to the project that someone wants, is it more valuable to give half of their stake to the other project in hopes of future collusion? If future collusion isn't a sure thing, that's probably not going to be a wise choice. Thus, changing out the group of delegates for Quadratic Voting substantially lowers the risk of this collusion.

\subsection{Simulation Results}

\begin{figure}[h!]
\centering
\includegraphics[width=\textwidth]{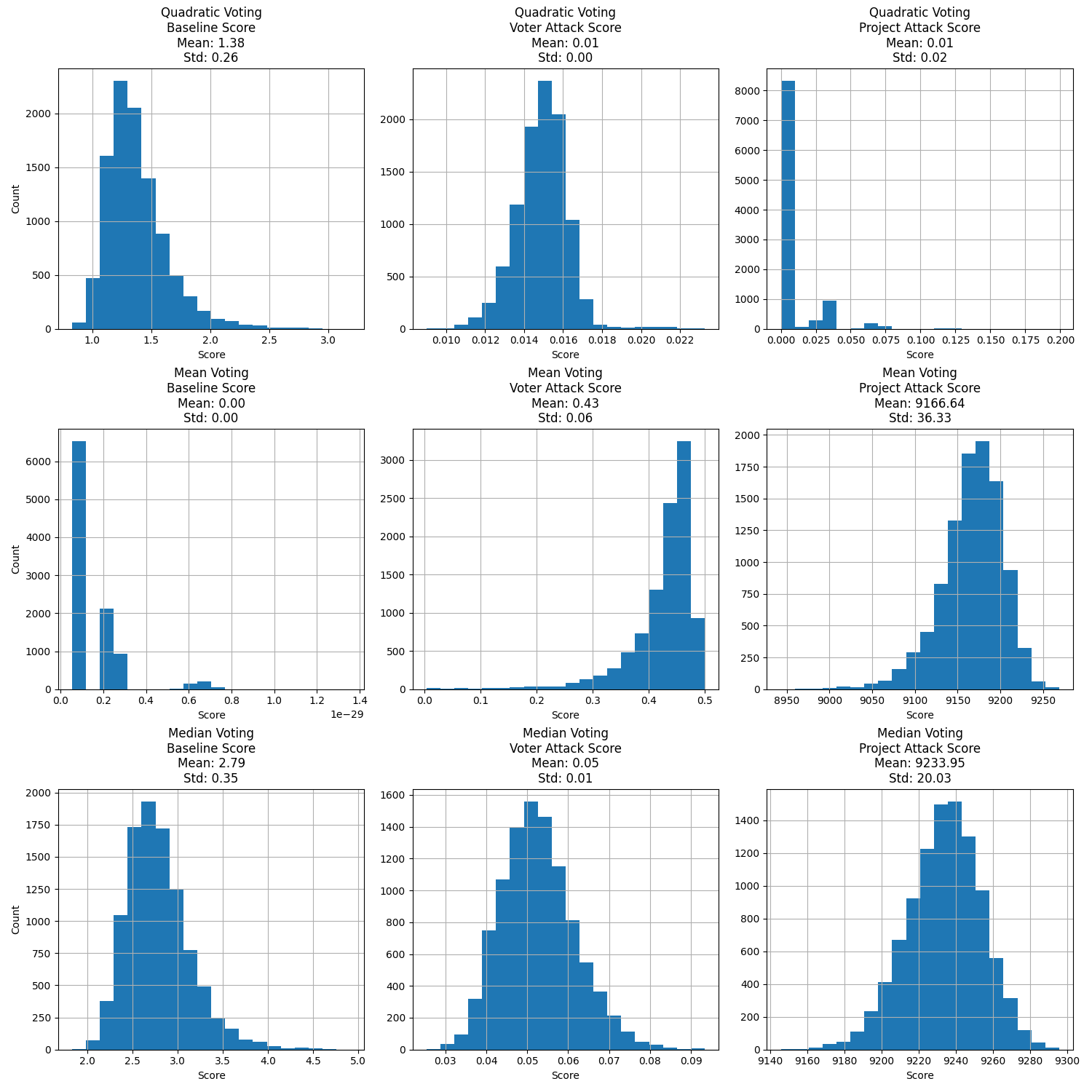}
\caption{Distribution of manipulation scores across voting mechanisms. Each row shows a different voting mechanism (Quadratic, Mean, and Median), while columns represent Baseline, Voter Attack, and Project Attack scenarios. The x-axis shows manipulation scores and y-axis shows frequency counts across 10,000 simulation iterations.}
\label{fig:voting_results}
\end{figure} 

Our simulation analysis substantiates the theoretical vulnerabilities proven in Section 4, while providing quantitative measures of each mechanism's resilience to attack. Figure \ref{fig:voting_results} presents the distribution of manipulation scores across baseline conditions and attack vectors for each mechanism. Drawing from the simulation results, we also highlight several observations:\\

\underline{Obversation 3: Quadratic Voting Demonstrates strong empirical resistance to manipulation} 

Quadratic voting demonstrates strong empirical resistance to manipulation. Under baseline conditions, the mechanism produces a normally distributed score centered at approximately $1.38$ with a standard deviation of $\approx 0.26$. More notably, voter attacks yield minimal manipulation scores in the $[0.013, 0.018]$ range, indicating robust defense against individual manipulation attempts. Project-level attacks show similar resilience, albeit with some outliers $>0.1$. These represent the smallest percentage differences relative to the other voting mechanisms.\\

\underline{Obversation 4: Both mean and median voting are highly vulnerable to project attacks}

Both mean and median voting mechanisms demonstrate severe vulnerability to project-level phantom attacks, with manipulation scores orders of magnitude larger than their baseline behaviors. Mean voting shows project attack scores concentrated in the $[9050, 9250]$ range, while median voting exhibits scores in the $[9140, 9300]$ range. These high scores align with our theoretical analysis of phantom vote attacks, where coordinated effort from multiple projects can significantly distort allocation outcomes. The normally distributed nature of these attack scores suggests that this vulnerability is systematic rather than probabilistic, indicating a fundamental weakness in both mechanisms against coordinated manipulation.\\

\underline{Obversation 5: Voter attacks affect mean more than median, project attacks vice versa}

The differential impact of attack vectors on mean and median voting reveals an interesting trade-off. Under voter attacks, mean voting shows manipulation scores clustering around $[0.4, 0.5]$, while median voting demonstrates more modest manipulation in the $[0.04, 0.07]$ range, showing approximately an order of magnitude difference. However, this relationship inverts under project attacks, where median voting shows marginally higher vulnerability (scores centered around $9233$) compared to mean voting (scores centered around $9166$). This empirically validates our theoretical analysis from Observation 1, suggesting that median voting suffers far more than mean in large-scale voting attacks, as a large number of colluding voters, as in the project attack case, will create a more significant cliff between ordered results. Conversely, when it comes to attacks involving individual voters, mean is an order of magnitude more susceptible to singular outliers than the median.

\subsection{Practical Recommendations}

\begin{table}[htbp]
\centering
\caption{Summary Evaluation of Voting Mechanisms}
\label{tab:voting-evaluation}
\begin{tabular}{|p{2.5cm}|p{11cm}|}
\hline
\textbf{Mechanism} & \textbf{Evaluation} \\
\hline
Quadratic Voting &

Very robust to both voter and project attacks compared to other mechanisms. To combat potential voter collusion, could be implemented with voter rotation periods, as well as with some minimum allocation thresholds.

\\
\hline
Mean Voting &

Simplicity in implementation and voter comprehension, no deviation in voter preferences vs. the baseline. But vulnerable to both voter attacks and project attacks. Slightly more resilient to large-scale project attacks than median.

\\
\hline
Median Voting &

Increased resistance to voter attacks vs. mean voting, but shows severe vulnerability to coordinated efforts due to the discrete step function. Would need complex mechanism re-engineering to be practically feasibility, but there is a big tradeoff for voter comprehension.

\\
\hline
\end{tabular}
\end{table}

Table \ref{tab:voting-evaluation} contains a summary of our practical recommendations derived from both our theoretical and simulation observations. To aid practitioners have access to our simulation framework, modify simulation parameters, and see how voting results may vary from a single set of preferences, we have released an open-source interactive prototype dashboard to demonstrate our results \footnote{A public prototype deployment with multiple customizable parameters can be found here: \url{https://opt-voting.vercel.app}}.

Cross-mechanism comparison reveals a clear ordering in manipulation resistance. Quadratic voting provides superior protection across all attack vectors, with manipulation scores consistently an order of magnitude lower than alternative mechanisms. Median voting offers moderate protection against individual voter attacks but remains vulnerable to coordinated efforts. Mean voting demonstrates the highest overall susceptibility to manipulation, particularly to project-level coordination. While quadratic voting shows inherent resilience, this protection assumes non-repeated interactions between voters, as discussed in Section 4.3. The stark contrast in project attack scores ($0.25$ vs $9200+$) between quadratic and other mechanisms emphasizes the practical importance of this theoretical distinction.

\section{Further Considerations}\label{sec:further}

\textbf{Median Voting Variations} For simplicity of theoretical and simulation demonstration, we modeled a standard median voting system. We note that within Round 3 and Round 4 of Optimism's RetroPGF mechanisms, Optimism modified the median voting mechanism to have quorum caps in Round 3 and maximum and minimum thresholds in Round 4 \cite{op_rpgf_r3} \cite{op_rpgf_r4}. While the intent of these mechanisms may have been to mitigate Phantom Vote style attacks, we believe that voter and project attacks remain plausible threats. In a quorum cap model for example, a group of $n$ attackers and a quorum of $q$ would need to simply calibrate an epsilon of $\epsilon > q/n$. Moreover, we hypothesize modified median mechanisms such as the min/max cap in RetroPGF Round 4 artificially collapse voter preferences, which may widely affect voter project choices, especially under our model that assumes preferences follow a Pareto distribution due to voter time and resource constraints. The artificial caps introduced onto the median seem conceptually very similar to quadratic voting, which is far more conceptually simpler to explain to voters. A comparison of these modified hybrid mechanisms with standard models of quadratic and median voting is an area for further research.\\

\textbf{Extending Simulation Parameters} Within our simulation framework, we made several model assumptions, including using a Pareto distribution for voter preferences, constants for voter weights, and using available project and voter data to approximate a realistic voting scenario. An area of further research is in evaluating this same set of voting mechanisms on different voter and project parameters, and using different distributions to model both voter preference and voter weight, empirically validate our assumptions against real-world, non-aggregated voter preference data. Our online simulation dashboard provides an extend toolkit to experiment with all of these additional parameters and with custom voter preference datasets.\\

\textbf{Conducting Voting Experiments} Our theoretical and simulation framework also lays the theoretical groundwork for future governance experiments. With quadratic voting being our tentative recommendation as a governance mechanism, we could experiment with implementing mechanisms for rotating voter sets for different projects to reduce collusion, while also balancing potential benefits of anti-collusion against the importance of maintaining expertise knowledge. This balance may be particularly crucial in the case of retroactive funding systems where there requires more sophisticated understanding of specific verticals and projects, and there may be a limited pool of informed voters. Another extension of this framework could be towards domains of decentralized voting outside of RetroPGF, such as in Optimism's Token House, to evaluate if similar mechanisms would be valuable there too.

\newpage
\bibliographystyle{plain}  
\bibliography{references}
\end{document}